\documentstyle[12pt]{article}
\begin{document} 

\title{Heavy Meson Production in Proton-Nucleus Reactions with Empirical
Spectral Functions\thanks{Supported by Forschungszentrum J\"ulich}}
\author{A. Sibirtsev, W. Cassing and U. Mosel \\
Institut f\"ur Theoretische Physik, Universit\"at Giessen \\
D-35392 Giessen, Germany}
\maketitle
\vspace{2cm}
PACS: 24.10.-i; 24.30.-v; 24.50.+g; 25.40.-h
\vspace{1cm}

\begin{abstract}
We study the production of $K^+, \rho, \omega$ and 
$\phi$ mesons in $p + ^{12}C$
reactions on the basis of empirical spectral functions. The high momentum,
high removal energy part of the spectral function is found to be negligible
in all cases close to the absolute threshold. Furthermore, the two-step
process ($pN \rightarrow \pi N N; \pi N \rightarrow 
N + \ K^+, \rho, \omega, \phi$)
dominates the cross section at threshold energies in line with earlier 
calculations based on the folding model. 
\end{abstract}

\newpage
 
\section{Introduction}
The production of heavy mesons in proton-nucleus reactions even below the
free nucleon-nucleon threshold is of specific
interest~\cite{1,2,Mosel,3,4} as one hopes to
learn about cooperative nuclear phenomena, high momentum components of
the nuclear wavefunction or in-medium modifications of the mesons
themselves. Furthermore, a precise knowledge of the meson production
channels in $p + A$ collisions is a necessary step towards a microscopic
understanding of meson production in nucleus-nucleus collisions, where
in-medium modifications should be enhanced due to the higher baryon
densities achieved~\cite{Bertsch5,Cassing3}.
Especially the production channels for the vector
mesons is of particular interest because their abundancy can be controlled
by dilepton spectroscopy independently. A possible 'softening' of the $\rho$
meson at high baryon density \cite{CSR} as possibly indicated by $e^+e^-$ and 
$\mu^+\mu^-$
data \cite{CERES,HELIOS,Ca,5} should also show up in the $\rho^0$ production 
on nuclei and could be controlled by the $\pi^+\pi^-$ invariant spectrum.
  
So far, the production of $\eta$ and $K^+$ mesons on nuclei has been
addressed, both 
theoretically~\cite{Cassing4,CugnonK,Golubeva,Mueller1,Buescher,Sibirtsev2} 
and 
experimentally~\cite{Koptev,Chiavassa1,Chiavassa2,Chiavassa3,Grosse1,Grosse2},
whereas only a few model studies on the production of $K^-$ and
$\rho, \omega, \phi$ mesons are available \cite{Golubeva,Sibirt}. 
Most of the former
approaches - though partly using the experimental momentum distribution for the
target of interest - employ a quasi-free dispersion relation for the 
struck target
nucleon and thus come in conflict with the A-body energy-momentum
conservation close to the absolute threshold because the struck target
nucleon is off-shell. 

Experimental information about the nucleon momentum {\it and} 
energy distribution is available from 
$(e, e^{\prime}p)$ reactions \cite{eep}. The data
show in addition to the 'mean-field' spectral function a high momentum,
high energy component which might be attributed to nucleon-nucleon
correlations with high relative but low center-of-mass momenta \cite{Atti1}.
The influence of this 'correlation' component on heavy meson production
in proton-nucleus collisions is of particular interest in our present study
as well as the role of primary and secondary (pion induced) reaction channels.

Our paper thus is organized as follows: In Section~2 we describe our approach
for meson production in proton induced reactions with spectral functions
and discuss its ingredients, i.e. the spectral function itself,
the elementary production cross sections
in $pN$ and $\pi N$ collisions, the effective number of collisions as well
as meson reabsorption effects in Section~3. By discarding possible 
in-medium effects we then present in Section~4 
our results for $p + ^{12}C$ reactions
for $K^+, \rho, \omega$ and $\phi$ mesons as a function of 
the beam energy $T_0$
with their channel decompositions. A summary and discussion of open problems 
concludes this paper in Section~5.     

\section{Reaction model for proton-nucleus collisions}

In terms of a multiple scattering theory the particle production in
proton-nucleus collisions can be described as an
incoherent sum of two-body interactions~\cite{Kerman,Newton,Jackson}. 
However, the individual interactions in nuclear matter might differ from
those in free space due to the average interaction of the collision
partners with the surrounding nuclear medium \cite{Satchler}.

The first term from the
multiple scattering theory
 we denote as the 'direct' or 'one-step' production
process. Furthermore, neglecting the medium effects on the reaction
partners, this approach corresponds to the
'impulse approximation' which can be applied whenever the relevant energy scale
is large compared to the nuclear binding energy.
Within the single-scattering approximation the 
Lorentz-invariant differential cross section
for meson production in $p+A$ collisions then can be written as
\begin{equation}
\label{moda}
E_M \frac{d^3\sigma_{pA\rightarrow
M X}}{d^3p_M} =
\frac {N_{eff}(E_M)} {A}
\int_{\Omega}  d^3 q \ d{\omega} \ S_P({\bf q},\omega) \ E_M^{\prime}
\frac{d^3 \sigma_{pN\rightarrow M X}(\sqrt{s})}
{d^3p_M^{\prime}}
\end{equation}
with $\sqrt{s}$ being the invariant energy  of  the  incident  
proton and the struck nucleon defined as
\begin{equation}
s = {\left( E_0^*  + \omega \right)}^2
-{\left( {\bf p_0^*} + {\bf q} \right)}^2
\end{equation}
where $E_0^*$, ${\bf p_0^*}$ and 
$\omega$, ${\bf q}$ are the total energy 
and the momentum of the beam and target nucleon, respectively.
Here the $^{*}$-indices denote the influence of the average nuclear optical
potential $U_{opt}$ which changes the kinetic energy of the incident 
proton as
\begin{equation}
T_0^* =T_0-U_{opt} ,
\end{equation}
where $T_0$ is the  beam kinetic energy and $E_0^*=T_0^*+m_N$, while
$m_N$ is the  nucleon mass. Note that the momentum of the incoming
proton ${\bf p}_0^{*}$ also has to be modified properly in 
case of vector potentials.
The momentum-dependent optical potential 
for cold nuclear matter at normal density  $\rho_0 \approx 0.158 fm^{-3}$
is about - 70 MeV for $p_0 = 0$,
becomes repulsive at beam energies above 300 MeV and reaches an almost 
constant value of 30 - 50 MeV \cite{Weber,Cooper,Hama,Welke}. Thus the optical 
potential for the impinging proton is the first medium effect that has to
be taken into account.

The  primed indices in Eq.~(\ref{moda}) 
denote the meson momentum and energy in 
the cms of the beam and the individual target nucleon, while
 $ E_M' d^3\sigma_{pN\rightarrow M X} /d^3p_M'$ is  the 
Lorentz invariant elementary   
cross section for meson production from $p+N$ collisions in free space.
The factor $N_{eff}$ accounts for  the effective primary collision  
number as well as for the in-medium absorption of the meson when 
propagating through the nuclear target as described
in Section~3.4. In  Eq.~(\ref{moda}) the index $\Omega$ stands for the Pauli
blocking of the final states.

The function $S_P(q, \omega)$\footnote{We adopt the normalization
$\int d^3q \ d\omega \ S_P(q,\omega) = A$, where $A$ denotes the target mass.}
 relates the momentum of the
target nucleon $q$ with its energy $\omega$. The free space relation,
${\omega}^2=q^2+m_N^2$, is no longer valid for a system of bound nucleons
because they are off-shell. In the mean-field approach the average
total energy of the bound nucleon is 
\cite{Sibirtsev2,Sibirtsev3}
\begin{equation}
\label{mfa}
\frac{1}{A} \sum_{j=1}^A {\omega}_j = \frac{M_A}{A} \simeq  (m_N -\epsilon )
\end{equation}
where $M_A$ denotes the mass of the target while
$\epsilon$ stands for the binding energy per nucleon.
Realistic spectral functions $S_P( q, \omega)$, as used in the present
calculations,  will be discussed in Section~3.1.

As proposed in \cite{Kilian,Cassing5} the dominant contribution to 
subthreshold heavy meson production should arise from a 
two-step reaction mechanism with an intermediate pion.
In accord with~\cite{Cassing4,Buescher} the
cross sections for $K^+, \rho$, $\omega$ and $\phi$-mesons from
the secondary pion induced reactions are calculated as
\begin{eqnarray}
\label{moda1}
E_M \frac {d^3 \sigma_{pA\rightarrow
M X}} {d^3 p_M} =
\frac {N_{eff}(E_M)} {A} \  \int_{\Omega}  d^3 q \ d{\omega}
\frac { d^3 p_{\pi }^{\prime} } {E_{\pi}^{\prime}}
\ S_P({\bf q}, \omega) \nonumber \\
\times E_M^{\prime\prime}
\frac {d^3 \sigma_{\pi N\rightarrow M X}(\sqrt{s})}
{d^3p_M^{\prime\prime} }
\ \frac {g_{\pi}  } { {\sigma}_{tot} } \times E_{\pi}^{\prime}
\frac { d^3 {\sigma}_{p N\rightarrow \pi X}(\sqrt{s^{\prime}}) }
{d^3 p_{\pi }^{\prime} } 
\end{eqnarray}
where the  double prime indices denote the system of the
intermediate pion and a target nucleon, while the single prime indices are
those for the beam and the first target nucleon. Moreover,
$ E_{\pi} d^3{\sigma}_{p N\rightarrow \pi X}/d^3p_{\pi}$
stands for the $\pi$-meson differential production 
cross section, which is calculated
in line with (\ref{moda}), while ${\sigma}_{tot}$ is the total
proton-nucleon cross section. We note that the on-shell elementary $\pi$-meson 
production e.g. can be described within the isobar model 
\cite{Cugnon1,Sibirtsev4}.
The factor $g_{\pi}$ in Eq. (\ref{moda1}) accounts for
the probability that the pion interacts again with a target nucleon 
(cf. Section 3.4).

\section{Ingredients of the model}

\subsection{The spectral function}
The in-medium relation between the energy $\omega$ and the
momentum ${\bf q}$ of the target nucleon relevant
to subthreshold particle production can be defined in analogy to
Guet and Prakash~\cite{Guet}, where
the energy $E_{prod}$ (available for the  meson production)
is related to the incoming  energy 
by total energy conservation as
\begin{equation}
E_0^{*} + M_A = E_{prod} +
{ \left( M_{A-1}^2 +{\bf q}^2 \right) }^{1/2} + E_{exc},
\end{equation}
where $M_A$ and $M_{A-1}$ are the masses of the initial and
residual nuclei, respectively, ${\bf q}$ is the internal momentum and
$E_{exc}$ stands for the excitation energy of the final nucleus.

The excitation energy $E_{exc}$ now is related to the removal
energy $E_R$ as~\cite{Frullani}
\begin{equation}
E_{exc} =  E_R +M_A -M_{A-1} -m_N ,
\end{equation}
where $E_R$ is the energy necessary to extract the
nucleon from the nucleus. Thus 
\begin{equation}
\label{rela}
\omega = m_N -  E_R
\end{equation}
with $E_R > 0$.   The probability to find a nucleon with momentum
${\bf q}$ and removal energy $E_R$ in a nucleus now is given by the 
nucleon spectral function $S(q, E_R)$. Accordingly,
the function $S_P(q, \omega)$ entering our calculations 
in~(\ref{moda}),(\ref{moda1}) can be evaluated from
the spectral function $S(q, E_R)$ by substituting Eq.~(\ref{rela}).

The  evaluation of the spectral function itself is quite an involved
problem~\cite{Benhar2,Mahaux,Atti} and still a matter of discussion.
In the present calculations we adopt $S(q, E_R)$ from 
Sick et al.~\cite{Sick} obtained within the orthogonal
correlated basis approach as well as a phenomenological spectral 
function from Ciofi degli Atti and Simula~\cite{Atti1}, that 
both provide a good 
description of $(e, e'p)$ reactions.

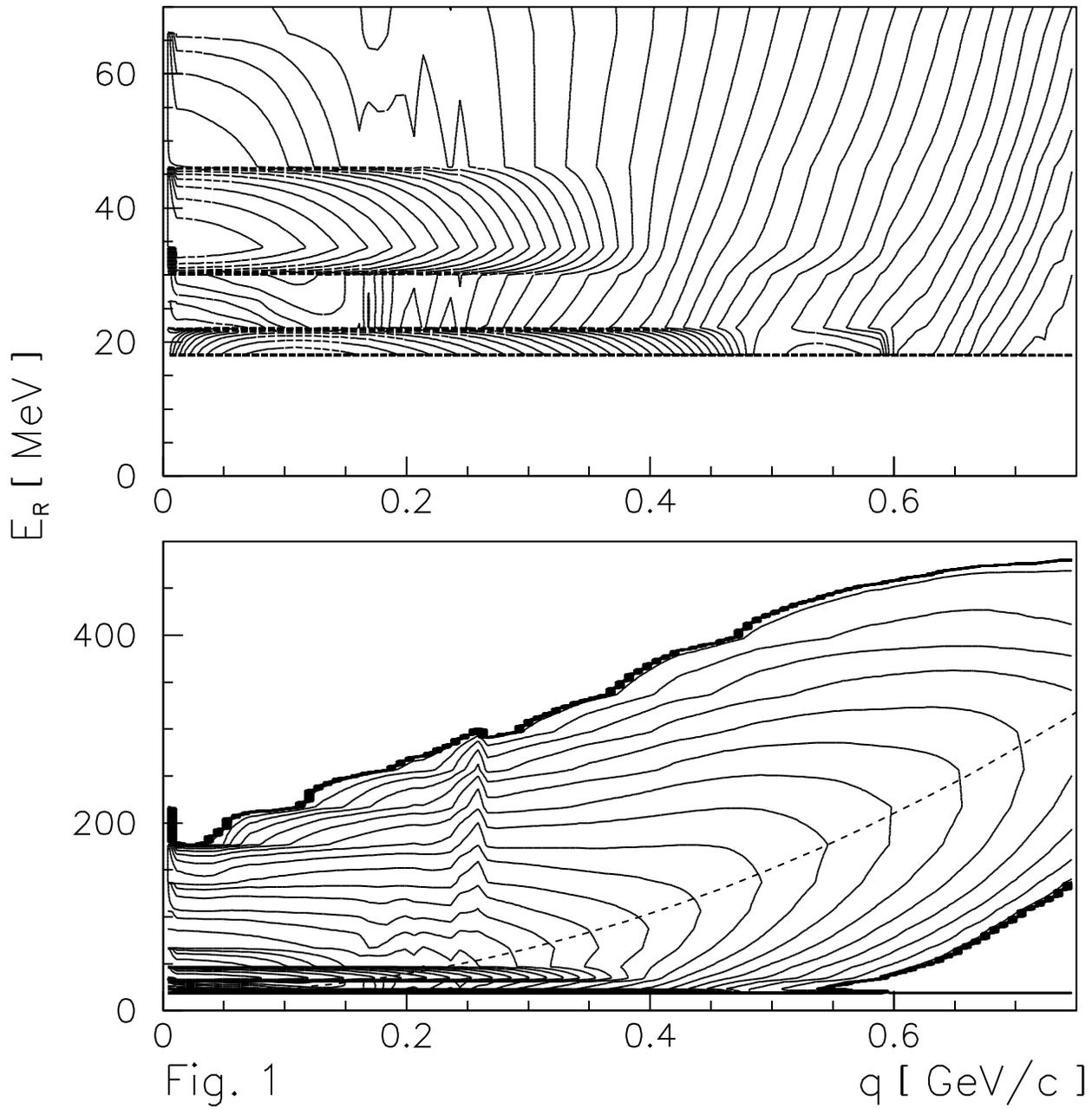
\begin{figure}
\caption{\label{s4}The spectral function $S(q, E_R)$ for
$^{12}C$ from~\protect\cite{Sick}. 
The dashed line shows the relation~(\protect\ref{cons}).
The contour lines decrease by a factor of 1.5 from
line to line.}
\end{figure}

The $S(q, E_R)$ for $^{12}C$ from ref.~\cite{Sick} is shown
in Fig.~\ref{s4} as a function of the nuclear nucleon momentum
$q$ and removal energy $E_R$. In the upper part of Fig. 2 we 
display the spectral function
up to $E_R \approx$ 60 MeV on a larger scale 
in order to show the 'mean-field' distribution
dominated  by the $s_{1/2}$ and $p_{3/2}$ single-particle
levels at $E_R \approx $ 35 MeV and $E_R \approx $ 20 MeV, respectively.
As a 'background' under the horizontal shell structure one can notice a ridge 
roughly along
\begin{equation}
E_R \simeq 45 MeV - \frac {q^2} {2 m_N}
\end{equation}
leading to a total available energy for particle production
\begin{equation}
\label{qfrm}
\omega \simeq m_N - 45 MeV + \frac {q^2} {2 m_N}.
\end{equation}
This is close to the dispersion relation often used in folding
model calculations.
 
The full spectral function up to $E_R =$ 500 MeV is given in the lower 
part where the
'mean-field' contribution only shows up as tiny ridges,
that extend up to $q\simeq 0.3$~GeV/c, i.e. up to the region of the
Fermi momentum. Beyond this value  the distribution
is dominated by the correlations.
The dashed line indicates the 
maximum of the spectral function for high momenta attributed to
short-range and tensor correlations as
\begin{equation}
\label{cons}
E_R = 2\epsilon + \frac{q^2} {2 m_N}
\end{equation}
with $\epsilon \simeq$7~MeV~\cite{Atti1}.
The simple relation (\ref{cons}) demonstrates that the high internal momenta
correspond to high removal energies, too. This is just opposite to the 
behaviour for momenta below the Fermi level and
will be quite important
for subthreshold particle production, because the contributions from 
the high momentum component of the nuclear wave function
should be negligible close to the absolute threshold since the 
available energy 
(including $E_R$) is below threshold.

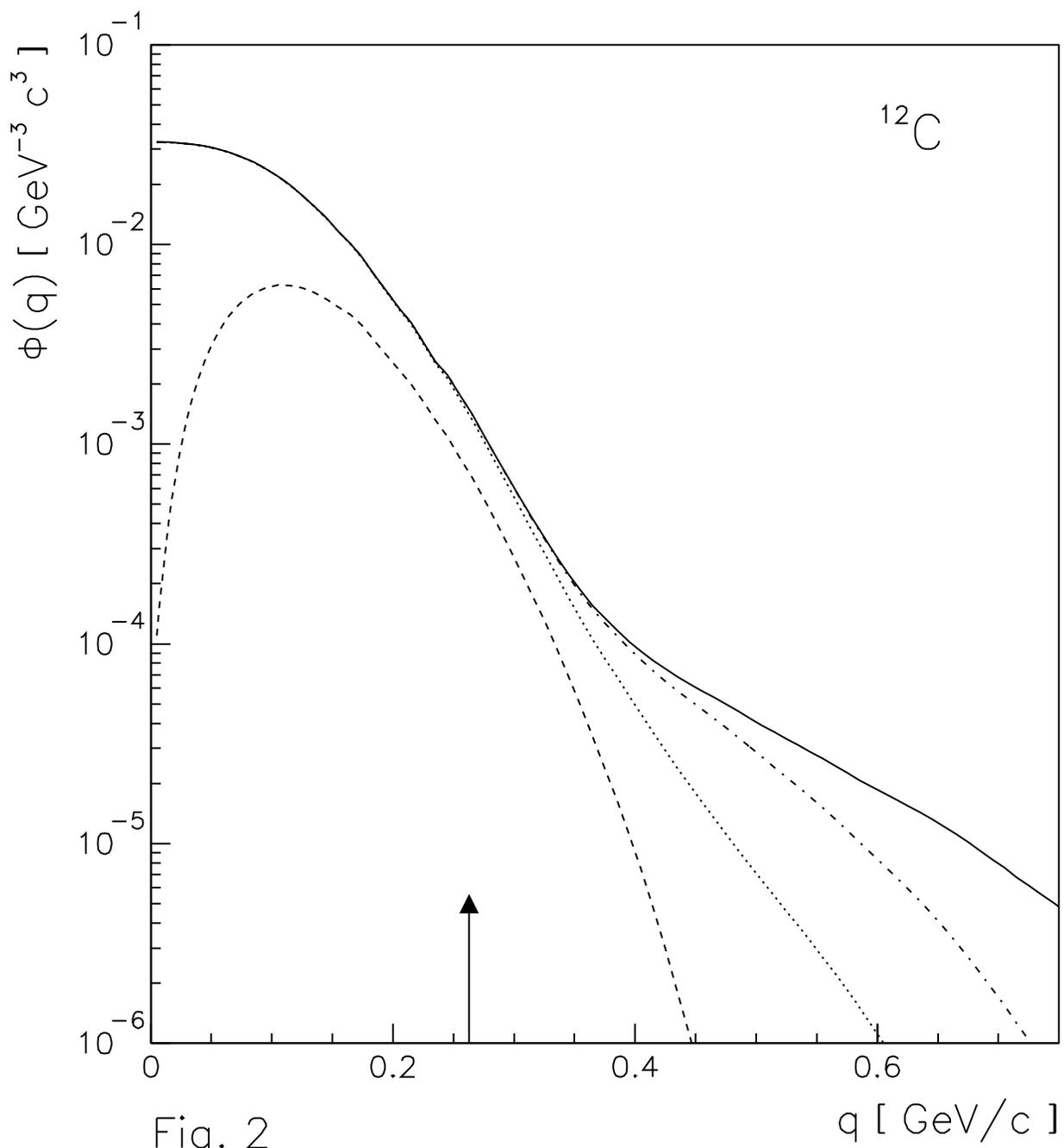
\begin{figure}
\caption{\label{s3}Saturation of the momentum sum rule for $^{12}C$.
The lines indicate the function~(\protect\ref{sat}) integrated up to
energy $E_{max}$=30 MeV (dashed), 100 MeV (dotted), 200 MeV (dash-dotted) and
500 MeV (solid line). The arrow indicates the Fermi momentum.}
\end{figure}

The relation between the high momentum and high removal energy
was for the first time illustrated by Ciofi degli Atti,
Pace and Salme~\cite{Atti} considering the saturation of
the momentum sum rule, i.e.
\begin{equation}
\label{sat}
\Phi(q)= \int_{E_{th}}^{E_{max}} dE_R \ S(q, E_R),
\end{equation}
where $E_{th}$ is the single nucleon removal threshold.
In Fig. \ref{s3} we show the function~(\ref{sat}) for a
$^{12}C$ target when integrating up to $E_{max}$ = 30, 100, 200 and 500
MeV. It becomes clear that momenta below the Fermi momentum are essentially 
related to a small removal energy, which is again in reasonable 
agreement with the mean-field approximation.

The function~(\ref{sat}), integrated up to infinity, is  
the nuclear momentum distribution. Respective momentum distributions for
$^3He, ^{12}C$, and $^{208}Pb$, 
that have been  calculated with the spectral functions from~\cite{Atti1}, 
are shown in Fig. \ref{s5}. We display separately
the contributions from the uncorrelated part 
(dashed line) as well
as the saturation of the momentum sum rule for the correlated part
at $E_{max}$=30 MeV (dotted line) and 100~MeV (dashed-dotted line). 
The uncorrelated part of the spectral function includes the ground and
one-hole states of the residual nucleus.
The short-range and tensor parts of the realistic $NN$ interactions
deplete the states below the Fermi level and partially occupy the
states above thus generating the
component of $S(q, E_R)$ with high momentum and high removal energy. 
The momentum distribution
corresponding to the correlated part for $E_{max}$ = 30 MeV (dotted lines)
is found to be small for all nuclei in comparison to the 'mean-field' 
contribution.

\subsection{Cross sections for pion induced reactions}

For our calculations we will need the elementary particle production 
channels over a wide kinematical regime and, in particular, down 
to the reaction threshold, where the experimental data often do not exist.
In this section we, therefore, formulate a parametrization that allows us
to perform an extrapolation of the cross section into the unmeasured region.

The cross sections for meson
production from $\pi+ N$ collisions can be separated
into two parts. The first one is an exclusive cross section related
to binary processes, which are dominant at energies close to the
reaction threshold, whereas the second part contains the
inclusive production at higher energies.

The cross section for the exclusive $K^+$ meson production from 
$\pi+ N$ collisions
has been calculated by Tsushima et al.~\cite{Tsushima}; we adopt their results
for our present study.

The cross section for $\rho$, $\omega$ and $\phi$-meson 
production from the reaction
\begin{equation}
\label{res1}
\pi +N \rightarrow R \rightarrow M+N
\end{equation}
is most conveniently  described within a resonance model. 
Assuming that the squared matrix element is proportional to
a Breit-Wigner function the cross section for the
reaction~(\ref{res1}) can be written as
\begin{equation}
\label{par2}
\sigma (\pi+N \rightarrow M+N; s) = \frac {\pi} {k^2} 
\frac {2J+1} {2}
\frac {B_{in} B_{out} {\Gamma}^2} {(\sqrt{s}-M_R)^2 + {\Gamma}^2/4}
\times R_2(s),
\end{equation}
where $J$, $B_{in}$ and $B_{out}$ are the resonance spin
and the branching ratios of the incoming and outgoing channels
respectively, while
the factor $R_2$ stands for the phase-space volume
of the final particles,
\begin{equation}
R_2=\pi \ {\lambda}^{1/2} \left( s, m_N, m_M \right)/s^{1/2},
\end{equation}
with $m_N$ and $m_M$ denoting the masses of the nucleon and 
meson, respectively.
In Eq.~(\ref{par2}) $k^2$ is given by
\begin{equation}
k^2=\lambda \left( s, m_{\pi}, m_N \right)
\end{equation}
with the K\"allen function 
\begin{equation}
\label{lam}
\lambda \left( z, x, y \right) = \left[ z- {\left( x+y
\right)}^2 \right] \left[ z-{\left( x-y \right)}^2 \right] / 4z.
\end{equation}
The parameters $M_R$ and $\Gamma$ from Eq.~(\ref{par2})
are the mass of the 'baryon resonance' and full width, 
respectively. We note that within the resonance
description we neglect coherent 
contributions from the available baryons as well as possible interference
terms since we do not have experimental information on their actual
magnitude.

Within our semiphenomenological approach we now fit the experimental
data on $\rho$, $\omega$ and $\phi$-meson production in pion induced 
reactions in order to extract the mass and width of the 
'effective' resonance $R$. 
Our fits to the data~\cite{Landolt} are shown in 
Figs.~\ref{pa4},\ref{pa5},\ref{pa6} with the parameters  
listed in Tab. \ref{ta1}. For convenience, the data are plotted as a 
function of 
$\sqrt{s}- \sqrt{s_{th}}$, with $s$ denoting the squared invariant energy 
of the colliding particles and $\sqrt{s_{th}}=2m_N + m_M $ in accordance with 
the reaction threshold. In Figs.~\ref{pa4},\ref{pa5},\ref{pa6} the fitted
resonance cross section for the exclusive reactions 
$\pi+ N \rightarrow \omega + N,
\rho+ N, \phi +N$, are given by the dashed lines in comparison to 
the data from \cite{Landolt}.

Since there are not enough experimental data available for 
the inclusive $\rho$, $\omega$ 
and $\phi$-meson production from $\pi + N$ collisions that allow
to construct reliable parametrizations, 
we calculate the cross
sections for the inclusive meson production within the 
Lund-String-Model (LSM) from~\cite{Lund}.
The calculated results within the Lund-String-Model 
can be fitted by a function of the form
\begin{equation}
\label{par1}
\sigma (\pi +p \rightarrow M+X) = a {\left( x-1 \right)}^b x^{-c}
\end{equation}
  from threshold up to a few 100 GeV,
where the scaling variable is defined as 
\begin{equation}
\label{scale}
x = s / s_{th}.
\end{equation}
The parameters for all interesting channels $a$, $b$, $c$
and $s_{th}$ are listed
in Tab.~\ref{ta2}.

Our fits to the  LSM results are shown
by the solid lines in Figs.~\ref{pa4},\ref{pa5},\ref{pa6} and
reasonably well reproduce the available inclusive data on vector
meson production from inclusive processes at high energies. For practical
purposes we will use the maximum of the 'resonance' cross 
section~(\ref{par2}) and the LSM parametrization~(\ref{par1}) for the
inclusive vector meson cross sections.

The Lorentz invariant differential cross sections entering Eqs.~(\ref{moda}) 
and~(\ref{moda1}),
finally, due to two-body kinematics near threshold are completely determined
when assuming an isotropic distribution in the pion-nucleon cms.  

\subsection{Proton induced reactions.}
The  cross sections for $K^+$, $\rho$, $\omega$ and $\phi$-meson
production from $p+p$ collisions within the Lund-String-Model (LSM) 
are shown in 
Figs.~\ref{kac1f},\ref{pa10},\ref{pa9},\ref{pa20} as a function of
the invariant collision energy $\sqrt{s}$ in comparison to the 
experimental data from~\cite{Landolt}. 
The solid lines show the fit to the LSM results
according to Eq.~(\ref{par1}) with the parameters from Tab.~\ref{ta4},
whereas in case of $\omega$ production the full dots stand 
for the LSM results.
We additionally show the cross sections for exclusive  meson 
production within the One-Boson Exchange 
Model (OBEM)~\cite{Sibirtsev1} by the dashed lines in comparison to the
respective data from \cite{Landolt} (triangles).
Whereas in case of $K^+$, $\rho$, $\omega$ production the inclusive LSM
results smoothly match with the OBEM calculations for the exclusive 
channels close to threshold, this is no longer the case for 
$\phi$-meson production (cf. Fig.~\ref{pa20}) because the string
fragmentation model requires the formation of two $s{\bar s}$-pairs,
which shifts the threshold up to higher energies. Since there are presently
no experimental data available for $\sqrt{s}<3.5$~GeV in case of $p+p$
reactions we will adopt the maximum of the cross sections from the 
OBEM and LSM calculations for the inclusive $\phi$-meson production 
as in the case of the $\pi +N$ reactions.

Though our present results for the $p+N$ channels have some uncertainty
close to threshold, this will not show up sensitively in $p +A$ reactions
since the pion induced channels contribute more effectively (cf. Section 4).

\subsection{The effective collision number}
The factor $N_{eff}(E)$ appearing in Eqs.~(\ref{moda}),~(\ref{moda1})
accounts for the $A$-dependence of the $K^+$, $\rho$, $\omega$ and
$\phi$-meson production as well as for the Final State
Interaction (FSI) of the mesons in the nuclear medium.
As first proposed by Margolis~\cite{Margolis1}, the $A$-dependence 
for particle production and propagation in the nuclear medium can be
described by extending the Glauber multiple scattering 
theory~\cite{Glauber}. In accordance with~\cite{Margolis2,Vercellin}
this gives
\begin{eqnarray}
\label{eq1}
N_{eff} & = &
\int_{0}^{+\infty} bdb \int_{-\infty}^{+\infty} \rho (b,z)dz
\int_{0}^{2\pi} d\phi \nonumber \\
&  & \times  \left[ exp \left( -{\sigma}_{pN} \int_{-\infty}^{z}
\rho (b,\xi ) d \xi - {\sigma}_{M N}
 \int_{0}^{+\infty}
\rho ({\bf r}[\zeta ]) d\zeta \right) \right],
\end{eqnarray}
where $\rho ({\bf r})$ is the single-particle density in coordinate space 
(taken from~\cite{Knoll}) 
normalized to the target mass number. Here,
${\bf r}[\zeta ]$  is 
\begin{equation}
{\bf r}[\zeta ] = {\bf r}_0(b,0,z)+ \zeta \, \hat{{\bf e}},
\end{equation}
where $b$ and $z$ stand for the impact parameter and the $z$-component of
the coordinate along the beam-axis,
respectively, and $\hat{{\bf e}}$ is a unit vector in
coordinate space in the form 
\begin{equation}
\hat{{\bf e}} :=
(sin \theta cos \phi , sin \theta sin \phi , cos \theta ),
\end{equation}
while $\theta $ is the meson emission angle in the
laboratory system. Moreover, ${\sigma}_{pN}$ is  the
total $p+N$ cross section while ${\sigma}_{M N}$ is the
$M+ N$ absorption cross section.

Within the small angle approximation, i.e. for $\theta \simeq 0^o$,
one can reduce~(\ref{eq1}) to
\begin{equation}
\label{eq2}
N_{eff}=
\frac {1} {{\sigma}_{pN}-{\sigma}_{MN}}
\int d^2b \ \left( exp \left[ -{\sigma}_{MN} \ T_A(b) \right] -
exp \left[ -{\sigma}_{pN} \ T_A(b) \right] \right) 
\end{equation}
with $T_A(b)$ denoting  the profile function of the target~\cite{Glauber}.
In Eq.~(\ref{eq2}) the second term is related to the effective number
of nucleons (of the target) participating in the interaction with the
impinging proton while the first term 
accounts for the final state reabsorption.

As shown in~\cite{Buescher,Cassing5} $N_{eff}$ from
Eq.~(\ref{eq2}) for ${\sigma}_{MN} \simeq 0$ can adequately
be parameterized by
\begin{equation}
\label{ALPHA}
N_{eff}=A^{0.75 \pm 0.01},
\end{equation}
at bombarding energies above 0.8 GeV,
where $A$ is again the target mass number. 
Thus the $A$-dependence of the direct production mechanism is
proportional to~(\ref{ALPHA}) whenever the produced meson only weakly 
interacts with the residual nucleus. In case 
of ${\sigma}_{pN} \gg {\sigma}_{MN}$  we can 
- assuming a uniform
distribution of the nuclear density - reduce the 
absorption term from~(\ref{eq2}) to the simple expression
\begin{equation}
\label{reab}
\kappa  = exp \left[ - {\sigma}_{MN}
{\rho}_0 R_A) \right],
\end{equation}
where ${\rho}_0 =0.168$ fm$^{-3}$ and $R_A$ are the average  density and
the radius of the nucleus, respectively. We note that 
the factor $\kappa $ is frequently used to describe the 
reabsorption of the produced particles in finite nuclei in order to 
account for weak reabsorption effects.

However, the application of~(\ref{reab}) in case of $\rho$ and $\omega$
production is not adequate because the reabsorption cross section
${\sigma}_{MN}$ appearing in~(\ref{eq1}-\ref{reab}) is no longer small
compared to $\sigma_{pN}$. In order to demonstrate the sensitivity of
$N_{eff}$ to the various approximation 
schemes~(\ref{eq1}),~(\ref{eq2}),~(\ref{reab}) we show
in Fig.\ref{coln} the effective collision number for $^{12}C$ as a function of
the reabsorption cross section within the limits of Eq.~(\ref{eq1}) 
(solid line), Eq.~(\ref{eq2}) (dashed line) 
and Eq.~(\ref{reab}) (dotted line) which deviate substantially
already for absorption cross sections of about 30 mb. Consequently, we
will use Eq.~(\ref{eq1}) in the following to describe the inclusive production
and absorption of mesons in $p + A$ reactions with the 
absorption cross sections described in Section~3.5

We, furthermore, note that by Eq.~(\ref{eq1}) we only account for reabsorption
and neglect elastic meson-nucleon rescattering. This is of no special
interest when considering inclusive (integrated) cross sections only as
in Section 4, but will be essential when comparing the Lorentz invariant
spectra with experimental data sets in a narrow kinematic regime. For
the latter purpose transport simulations as described in ref. \cite{Rudy}
will provide a more adequate description. 

The calculation of $N_{eff}$ for the two-step reactions with an intermediate
pion has been described by Vercellin et al.~\cite{Vercellin}
within the Glauber formalism and is a more involved task. However, as 
shown in~\cite{Cassing5},  a practicable approximation is to
use $N_{eff}$ from Eq.~(\ref{eq1}) and to include the corrections from the
two-step mechanism by a factor
$g_{\pi}$ as
\begin{equation}
g_{\pi} = 1- \frac{1}{4 R_A^2}  {\left(
\frac {1} { {\sigma_{pN} \rho_0}} + \frac {1} { {\sigma_{\pi N}\rho_0}}
\right)}^2,   
\end{equation}
which describes the average probability for a pion to scatter again in the
target with radius $R_A$. In case of small final meson ($K^+$, $\rho$,
$\omega $, $\phi$) reabsorption 
the $A$-dependence of the inclusive production cross section 
due to the reactions
with an intermediate pion roughly scales as  
\begin{equation}
N_{eff} \times g_{\pi} \simeq A^{1.2 \pm 0.1} .
\end{equation}

\subsection{$\rho, \omega, \phi$-nucleon cross sections}
The total cross sections for $V+N$ interactions at high energies can be
evaluated with the Vector-meson Dominance Model (VDM) from $\rho$,
$\omega $ and $\phi$-meson photoproduction~\cite{Sakurai}
or be obtained within the additive quark model
relations~\cite{Joos,Kajantie} from measurable cross sections.
The results obtained with the VDM~\cite{Christillin} are in good
agreement with those computed via quark sum rules, resulting in
${\sigma}_{\rho N} = {\sigma}_{\omega N} \simeq 25$ mb and
${\sigma}_{\phi N} \simeq 12$ mb at vector meson momenta
above 1 GeV/c.

The cross sections for $\rho +N$, $\omega +N$ and $\phi +N$
interactions in case of low relative momenta can be extracted from the 
$\pi + N$ reactions 
by assuming that this is the dominant absorption channel.
Detailed balance provides the relation
\begin{equation}
\label{bal}
\frac {{\sigma}_{\pi +N \rightarrow M+N}} 
{{\sigma}_{M +N \rightarrow \pi +N}} = \frac {(2I_N+1) (2I_M+1)}
{(2I_{\pi}+1) (2I_N+1)} \frac {\lambda (s,m_N,m_M)}
{\lambda (s,m_N,m_{\pi})} ,
\end{equation}
where $s$ is the squared invariant energy, $m_N$, $m_{\pi}$,
$m_M$ are the masses, while $I_N$, $I_{\pi}$ and $I_M$ 
are the spins of nucleon, pion and  vector meson, respectively. 
Here, the function $\lambda$ is given by (\ref{lam}).
With our parametrizations for the cross sections
${\sigma}_{\pi +N \rightarrow M+N}$~(\ref{par2}) we then obtain the 
cross sections for vector meson-nucleon interactions via~(\ref{bal}).

The cross sections extracted from detailed balance
are only related to the particular inelastic reaction channel 
$M N \rightarrow \pi N$, and do not saturate the total cross sections
even at  momenta of 100 MeV/c.
Indeed the $2 \pi N$, $3 \pi N$, $m\pi N$ or $K \bar{K} N$ final states
as well as elastic scattering can also occur,
which all contribute to the total cross sections. Since there are no
reliable calculations on the $V+N$ cross sections at
low energy~\cite{SibirtsevD},  we adopt the results from the VDM.

\section{Results for $p + ^{12}C$}
After specifying all the elementary cross sections in proton and pion
induced channels as well as the spectral function $S(q,E_R)$ 
\cite{Atti1,Sick} we now
present the results for meson production cross sections 
in $p + ^{12}C$ collisions according 
to ~(\ref{moda}) and~(\ref{moda1}). Though our primary 
interest are the vector mesons,
we first discuss our calculations for $K^+$ production in comparison to
the data from Koptev et al.~\cite{Koptev} 
and the results from the folding model~\cite{Cassing4}. 
In Fig.~\ref{e6ka} we thus compare 
the cross sections within the spectral
function approach for the one-step channel (dashed line) and the two-step
mechanism (solid line) with the one-step (dash dotted line) and two-step
result (dotted line) from the folding model with elementary cross sections
as described in ref. \cite{Cassing5}. 

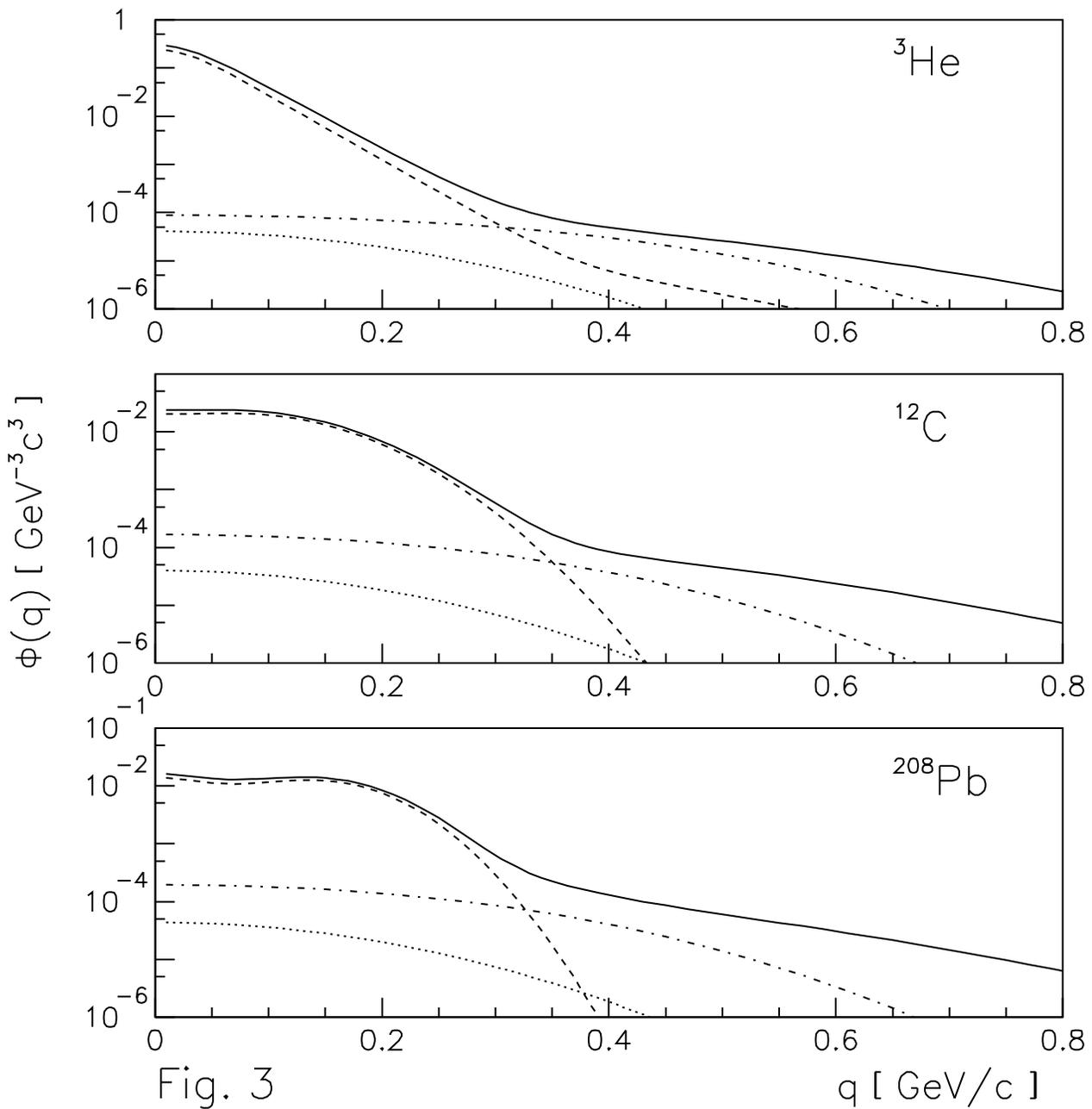
\begin{figure}
\caption{\label{s5}The momentum distribution for $^3He$, $^{12}C$,
and $^{208}Pb$ according to~\protect\cite{Atti1}.
The dashed lines result from the uncorrelated part of the
spectral function. The dotted lines
indicate~(\protect\ref{sat}) for the correlated spectral function
integrated up to $E_{max}=30$ MeV and the dashed-dotted lines 
up to $E_{max}=100$ MeV.  The solid lines display the sum
of the uncorrelated and correlated parts for 
$E_{max} \rightarrow \infty $.}
\end{figure}

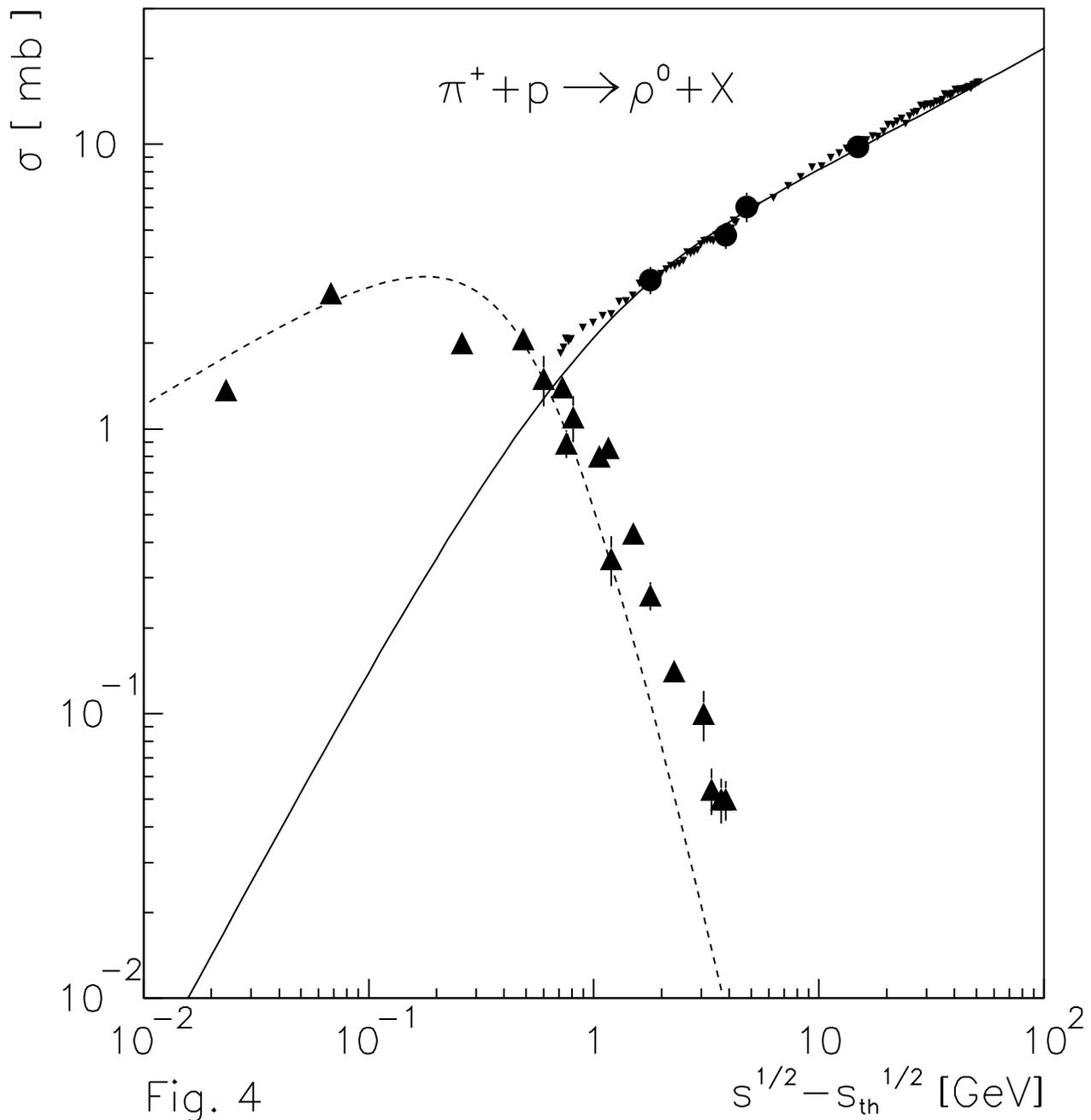
\begin{figure}
\caption{\label{pa4}Experimental  cross sections
for the exclusive reaction $\pi^++p \rightarrow \rho^++p$ (triangles)
and the inclusive reaction 
$\pi^++p \rightarrow \rho^o+X$ (full dots) from~\protect\cite{Landolt}. 
The lines represent our parametrizations as
discussed in the text.}
\end{figure}

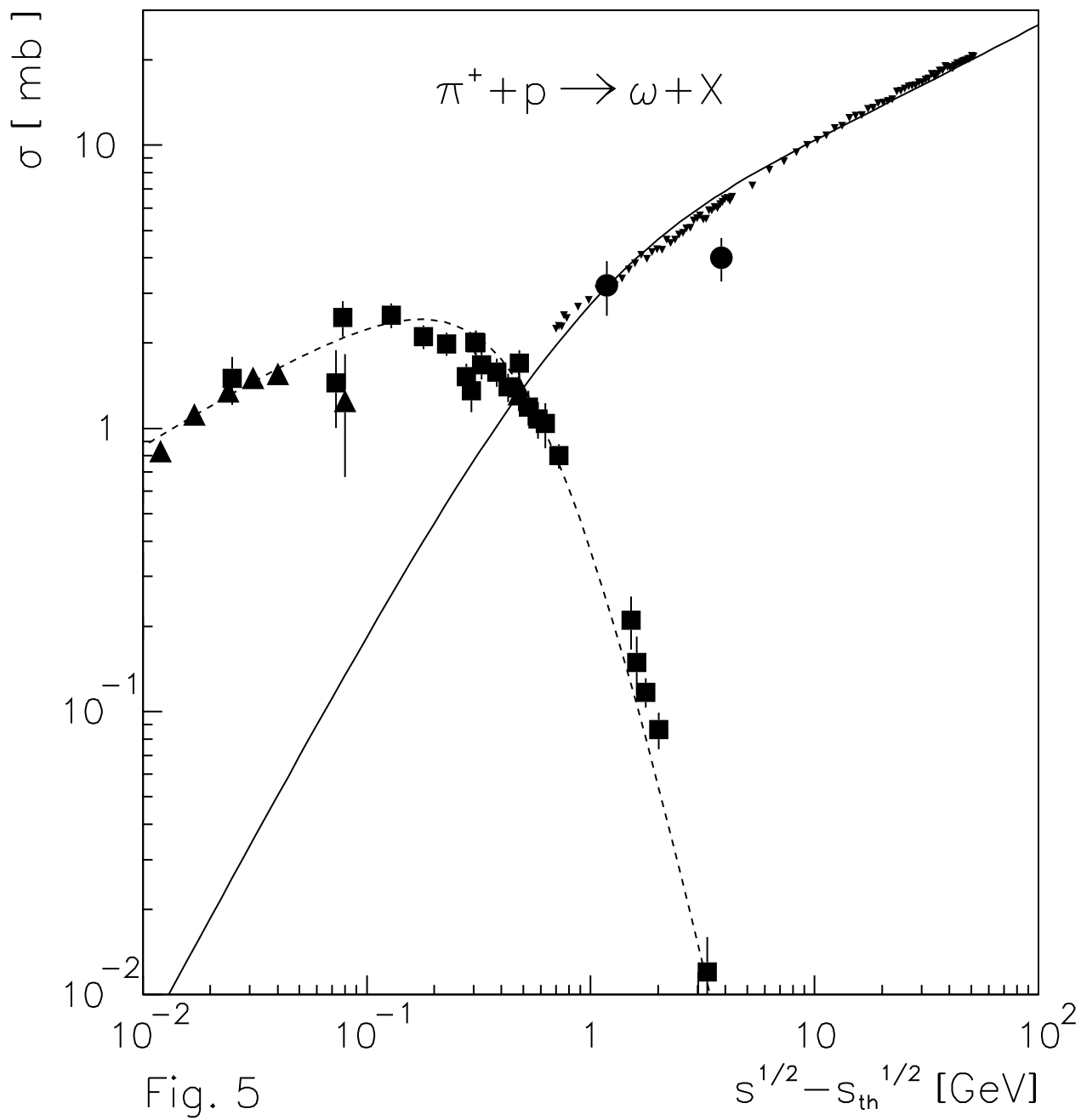
\begin{figure}
\caption{\label{pa5}Experimental  cross sections
for the exclusive reactions $\pi^++n \rightarrow \omega+p$ (squares) 
and  $\pi^-+p \rightarrow \omega+n$ (triangles) and the inclusive reaction
$\pi^++p \rightarrow \omega+X$ (full dots). 
Experimental data are taken from~\protect\cite{Landolt}. 
The lines represent our parametrizations as
discussed in the text.}
\end{figure}

Note that
the folding model, contrary to the spectral function approach, 
is a pure mean-field approximation, where the nucleon momentum distribution for
the target nucleus is calculated in the Hartree-Fock limit~\cite{Cassing5}
corresponding to the dashed line in Fig.~\ref{s5}.
Thus the high momentum components, that are associated with the 
correlated part of the spectral function (cf. Fig.~3), are missing in
the folding model calculations~\cite{Cassing5}. 
The dispersion relation for a target nucleon in the folding model was
taken as 
\begin{equation}
\label{disr}
\omega (q) = \sqrt{ q^2+ m_N^2} - \epsilon
\end{equation}
where $\epsilon $ is again the average binding energy per nucleon.
In spite of the quite different assumptions in
both approaches, the results for the $K^+$-meson cross sections for
one-step and two-step reaction channels are remarkably close to
each other.

Whereas at energies above 0.95 GeV both results are practically
the same, the folding model with the quasi-free dispersion 
relation~(\ref{disr}) for the
target nucleon overestimates the energy available in the collision and
thus leads to slightly higher cross sections than our spectral function
approach for both reaction steps at lower bombarding energy. 
Thus the underestimation of the data
below about $T_0 =0.92$ GeV  might be considered as an
evidence for additional many-body effects or further production channels
that become important close to the reaction threshold. On the other hand,
the dominance of the two-step reaction mechanism is the same within both
approaches and can be considered as model independent. 

Since in the spectral function from~\cite{Sick} we cannot separate the 
correlated and uncorrelated parts, we adopt the 
spectral function from~\cite{Atti1}, that allows the separation 
of the two components.
Within the latter parametrization we can study the relative 
contribution from the correlated part of $S_P(q,w)$ for meson 
production in $p+A$ reactions.
Our calculated results for $K^+$-meson production in $p+^{12}C$
collisions are shown in Fig.~\ref{e6ka1} for the two-step
mechanism (solid line) and primary reaction channel (dashed line)
in comparison with the experimental data~\cite{Koptev}.
In line with our arguments presented above the contributions from
the correlated part of the spectral function indeed are small
compared to the data.

For more pedagogical reason we also discuss the results for the $K^+$-meson
production cross section when employing the momentum distribution 
$\Phi (q)$, as obtained from integrating the spectral 
function from~\cite{Atti1} over the removal energy, but using the 
free dispersion relation 
\begin{equation}
{\omega}=\sqrt{ m_N^2 + {\bf q}^2},
\end{equation}
for all momenta with $m_N$ being the mass of free nucleon
and ${\bf q}$ the nuclear momentum taken in accordance with the
function ${\Phi}(q)$ from~\cite{Atti1} shown in Fig.~\ref{s5}.

In this unrealistic limit the one-step production mechanism 
(dashed-dotted line in Fig.~13) still underestimates the experimental
data by more than a factor of 5, however, the results for the two-step
mechanism now overestimates the data by almost an order of magnitude.
This comparison clearly demonstrates that the high momentum component 
appearing in the spectral function approach cannot be
exploited in a quasi-free production mechanism.

We continue with our calculations for the vector mesons in 
$p + ^{12}C$ reactions. The results
are shown in  Fig.~\ref{e6h}; here, the solid
lines indicate the contribution from the two-step reactions 
with an intermediate
pion (\ref{moda1}), while the dashed lines correspond to the direct
production mechanism (\ref{moda}) using again the 
spectral function from~\cite{Sick}. For all mesons ($\rho $,
$\omega $, $\phi $) the two-step mechanism is clearly dominant
close to threshold 
and kinematically is due to the effect that the Fermi motion of the nucleons
is exploited twice in the two-step channels. The absolute cross sections
at about 2 GeV are in the order of 10 $\mu$b for 
$\rho $, $\omega $-production, which is high enough to allow for
a detailed experimental study at COSY or CELSIUS.

\section{Summary}
In this work we have presented an approach for meson production in p + A
reactions on the basis of direct and two-step production channels (including
an intermediate pion) on the basis of the nuclear spectral function without
incorporating any selfenergies for the mesons. The 'elementary' cross sections
for $\rho, \omega$ and $\phi$ production have been evaluated within the
LUND string formation and fragmentation model~\cite{Lund} 
and within a resonance model
for $\pi N$ collisions close to threshold, respectively. 
The available
experimental data from~\cite{Landolt} for the exclusive and 
inclusive reactions
are described sufficiently well, however, our extrapolations for the various
channels close to threshold should be controlled experimentally in the near
future.

Whereas in previous reaction models only the target momentum distribution
has been employed using a quasi-free dispersion relation for 
the target nucleons,
our present formulation is based on the full spectral function $S(q,\omega)$
where the nucleon may be far off-shell and high momentum, high removal
energy components appear as well. 

This approach was first used by 
Debowski, Grosse and Senger~\cite{Grosse1,Grosse2} for proton 
induced particle production. While those authors applied the
method to their own measured differential spectra for
$K^+$-meson production on $^{12}C$ and $^{208}Pb$, we have 
tested our approach for $K^+$ production in comparison
to the experimental data from Koptev et al.~\cite{Koptev}
using the spectral function for $^{12}C$ from
ref.~\cite{Sick}.

What emerges from a detailed study of the spectral function is that 
one should use a quasifree dispersion relation~(\ref{qfrm}) only
for nuclear momenta below  the Fermi level. For higher values of $q$ a 
dispersion relation  with~(\ref{cons}) should be used
for the origin of the high momentum component accounting
for two-body correlations.

It is clearly seen
that the two-step mechanism with an intermediate pion gives by far the
dominant contribution at subthreshold energies as in the 
folding model~\cite{Cassing4}, but
the actual cross section for $K^+$-meson
is underestimated below about 920 MeV indicating
the need for further reaction channels or many-body effects. This result
contrasts with the folding model where a better description of the data is
achieved. We attribute this to the fact that the folding model - adopting
a quasi-free dispersion relation for the target nucleon - overestimates the 
available energy in nucleon-nucleon collisions when approaching the
absolute reaction threshold.

On the other hand, when integrating the spectral function over the energy
$E_R$, we obtain a momentum distribution showing very high momentum
components that can be attributed to short-range nucleon-nucleon or
tensor correlations. However, these high momentum components on average
are associated with high removal energies, too, such that they cannot be
exploited for meson production close to the absolute reaction threshold.
Thus, meson production close to threshold does not provide information
on the high momentum components in finite nuclei or short-range two-body
correlations.

Within our approach we have also calculated the cross sections for the vector
mesons $\rho $, $\omega $, $\phi $ in $p + ^{12}C$
collisions from threshold up
to a few GeV. We  obtain estimates for their production in nuclei without
incorporating any selfenergies for the mesons. These cross sections provide
a baseline for experiments at COSY or CELSIUS that address the in-medium
properties of the vector mesons as well as their final state interaction
with nuclei. Here, in context with chiral symmetry
restoration, the $\rho$ and $\omega$-mass at normal nuclear density $\rho_0$
is expected
to be reduced by about 18\% \cite{CSR}, which will sizeably enhance their
production cross section as compared to our calculations for free vector
mesons presented above. A detailed study of their production cross section
as a function of $T_0$ and target mass $A$ might help to separate the
effects from the real and imaginary part of the meson selfenergies.

\vspace{2cm}
The authors acknowledge many fruitful discussions with M.~B\"uscher,
E.~Grosse and 
B.~Kamys. They like to thank, furthermore, I. Sick for providing the $^{12}C$ 
spectral function in numerical form.

\newpage
\begin{table*}[h]
\begin{center}
\caption{\label{ta1}The  effective mass and
width of the baryonic resonance in Eq.~(\protect\ref{par2}),
as well as  $B=B_{in}\times B_{out}\times (2J+1)$.}
\vspace{0.6cm}
\begin{tabular}{|l|c|c|c|}
\hline
Meson & $M_R$ (GeV) & $\Gamma $ (GeV) & $ B$ ($\mu$bGeV$^{-2}$)\\
\hline
${\rho}^0$ & 1.809 & 0.99 & 413 \\
$\omega $ & 1.809 & 0.99 & 302 \\
${\phi} $ & 1.8 & 0.99 & 5.88 \\
\hline 
\end{tabular}
\end{center}
\end{table*}

\begin{table*}[h]
\begin{center}
\caption{\label{ta2}The parameters of Eq. (\protect\ref{par1})
for  pion induced reactions. Here the index $^{(1)}$ implies a shift 
$x \rightarrow x-1.3$. }
\vspace{0.6cm}
\begin{tabular}{|l|c|c|c|c|}
\hline
Meson & $s_{th} $ (GeV$^2$) & $a$ (mb) & $b$ & $c$ \\
\hline
${\rho}^0$ & 2.917 & 3.6 & 1.47 & 1.25 \\
$\omega $ & 2.958 & 4.8 & 1.47 & 1.26 \\
${\phi}^{ (1)}$ & 3.831 & 0.09 & 2.54 & 2.1 \\
\hline 
\end{tabular}
\end{center}
\end{table*}

\begin{table*}[h]
\begin{center}
\caption{\label{ta4}The parameters of Eq.~(\protect\ref{par1})
for  proton induced reactions. Here the index $^{(1)}$ implies 
a shift $x \rightarrow x-1.3$. }
\vspace{0.6cm}
\begin{tabular}{|l|c|c|c|c|}
\hline
Meson & $s_{th} $ (GeV$^2$) & $a$ (mb) & $b$ & $c$ \\
\hline
$K^+$ & 6.49 & 1.12 & 1.47 & 1.22 \\
${\rho}^0$ & 7.01 & 2.2 & 1.47 & 1.1 \\
$\omega $ & 7.06 & 2.5 & 1.47 & 1.11 \\
${\phi}^{ (1)}$ & 8.38 & 0.09 & 2.54 & 2.09 \\
$\eta $ & 5.88 & 2.5 & 1.47 & 1.25 \\
\hline 
\end{tabular}
\end{center}
\end{table*}
\newpage

\begin{figure}
\caption{\label{pa6}Experimental  cross sections
for the exclusive reactions $\pi^++n \rightarrow \phi+p$ (squares) and  
$\pi^-+p \rightarrow \phi +n$ (triangles) and inclusive reaction
$\pi^++p \rightarrow \phi +X$ (dots) from~\protect\cite{Landolt}.
The lines represent our parametrizations as
discussed in the text.}
\end{figure}

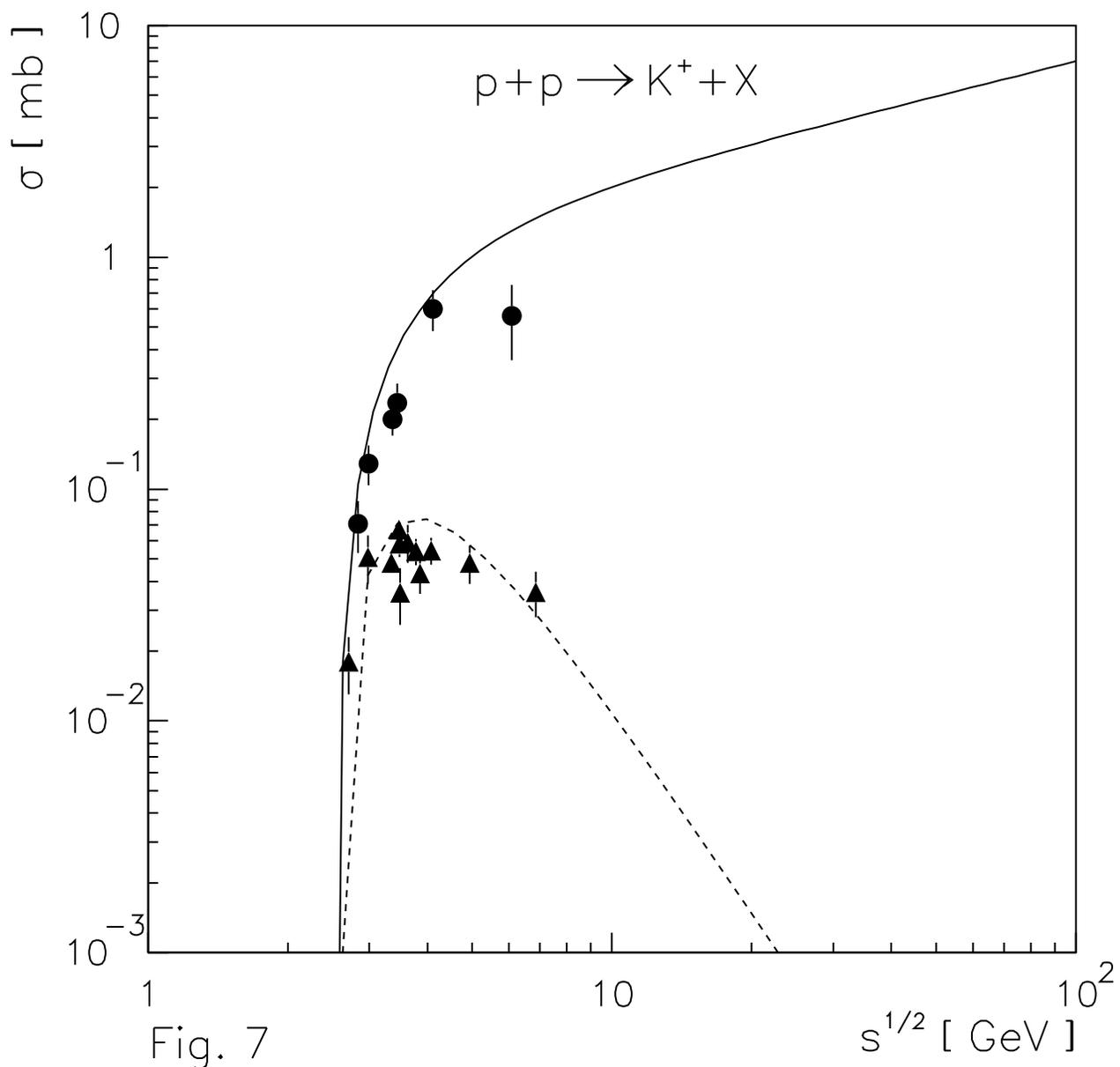
\begin{figure}
\caption{\label{kac1f}Cross sections
for the  exclusive reaction $p+p \rightarrow K^+ +\Lambda +p$ (triangles) and 
the inclusive reaction
$p+p \rightarrow K^+ +X$ (dots) from~\protect\cite{Landolt}.
The solid line shows the calculation within the LSM, while the dashed 
line refers to the OBEM.}
\end{figure}

\begin{figure}
\caption{\label{pa10}Cross sections
for the exclusive reaction $p+p \rightarrow \rho^o+p+p$ (triangles) and 
the inclusive reaction
$p+p \rightarrow \rho^o +X$ (dots) from~\protect\cite{Landolt}.
The solid line shows the calculation within the LSM, while the dashed 
line refers to the OBEM.}
\end{figure}

\begin{figure}
\caption{\label{pa9}Experimental cross sections
for the exclusive reaction $p+p \rightarrow \omega+p+p$ (triangles)  
  from~\protect\cite{Landolt} in comparison to the OBEM 
calculation (dashed line). The full dots represent the results from LSM for
the inclusive reaction $p+p \rightarrow \omega+X$, while the solid line 
shows our fit with Eq.~16.}
\end{figure}

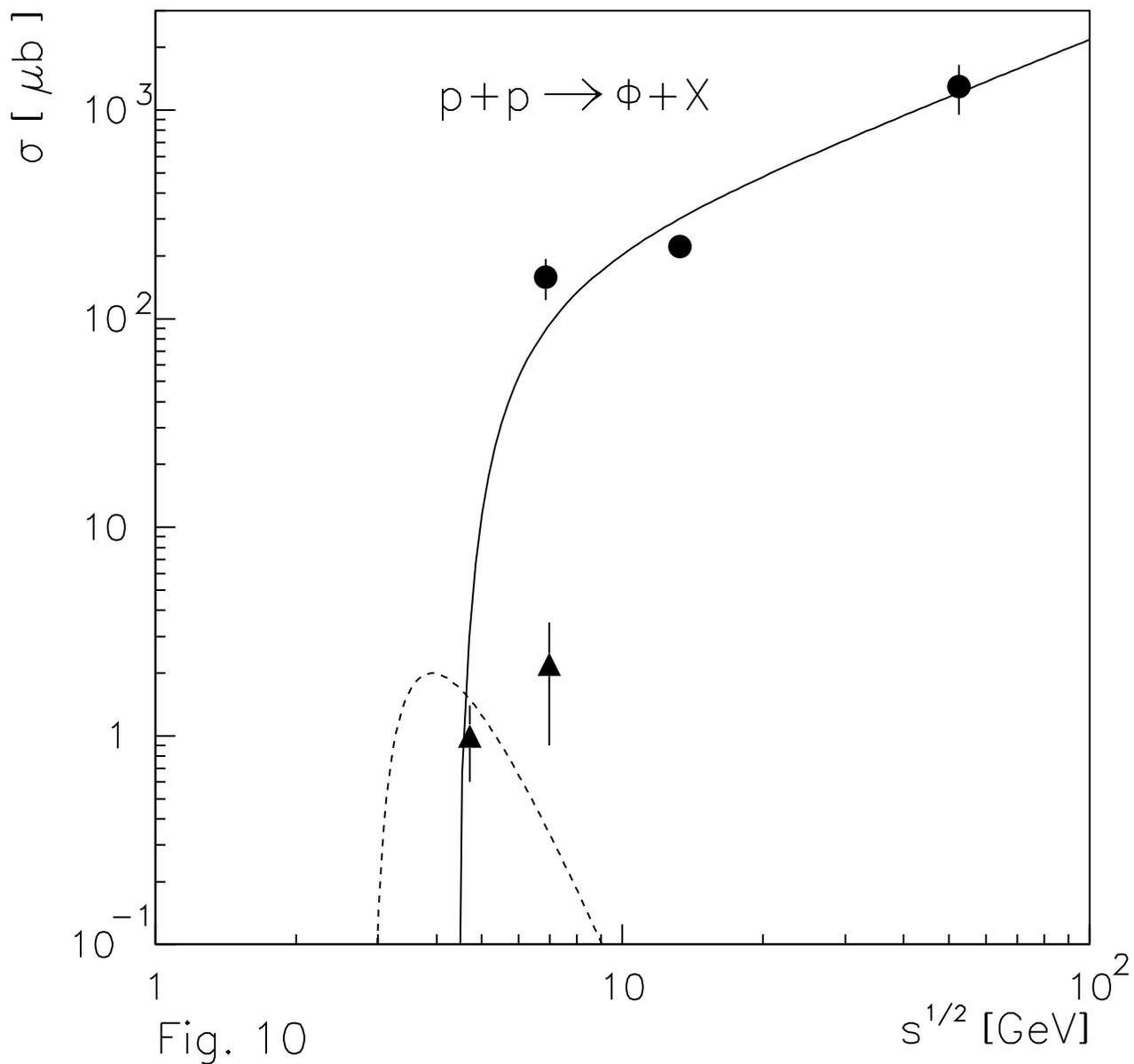
\begin{figure}
\caption{\label{pa20}Cross sections
for the  exclusive reaction $p+p \rightarrow \phi+p+p$ (triangles) and
the inclusive reaction
$p+p \rightarrow \phi +X$ (dots) from~\protect\cite{Landolt}. The
solid line shows the calculation within the  LSM and the dashed line 
our results within the OBEM for the exclusive channel.}
\end{figure}

\begin{figure}
\caption{\label{coln}The effective collision number calculated 
with~(\protect\ref{eq1})-solid, (\protect\ref{eq2})-dashed
and with (\protect\ref{reab})-dotted line as a function of the
meson absorption cross section $\sigma $ in mb.}
\end{figure}

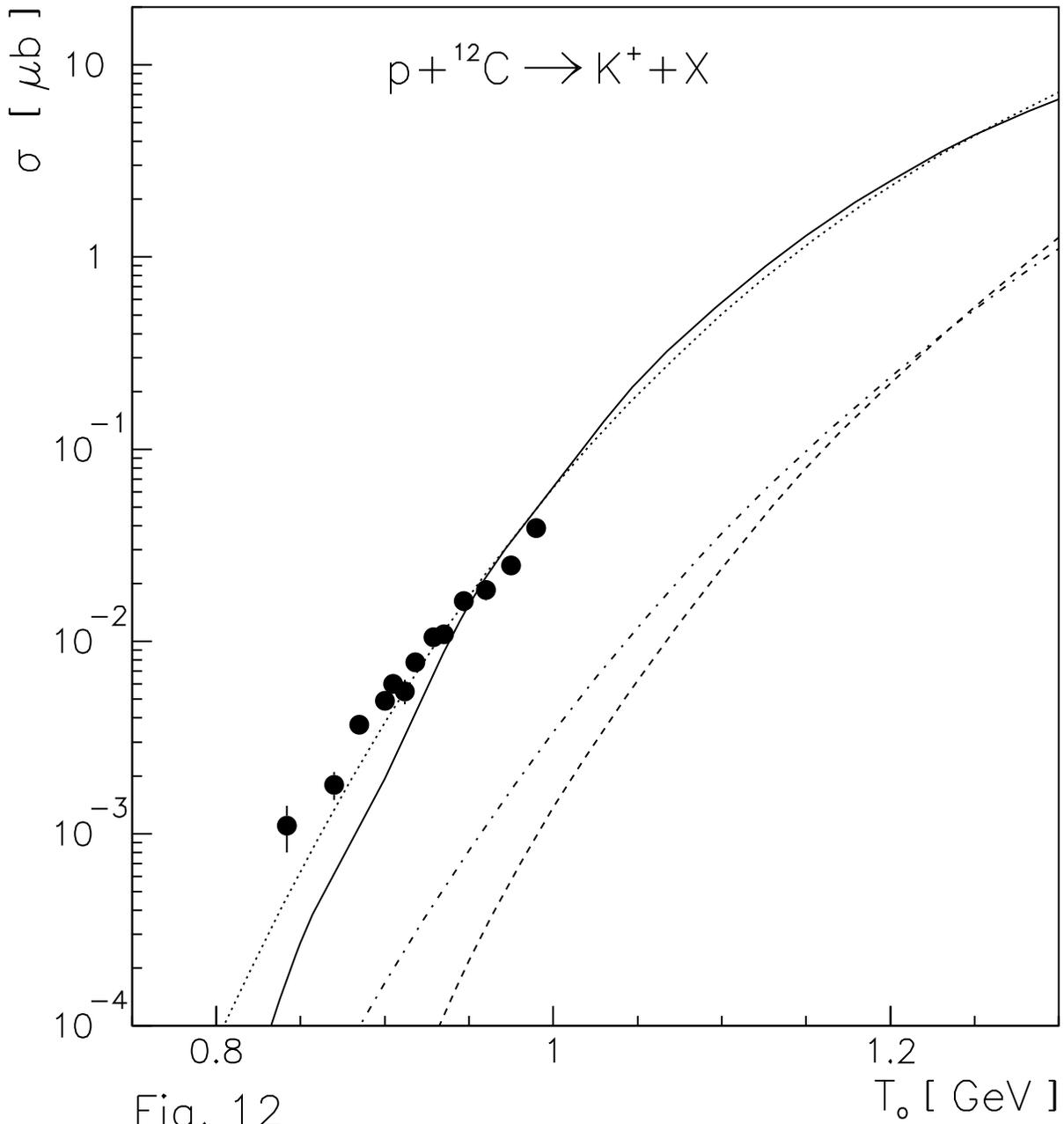
\begin{figure}
\caption{\label{e6ka}Cross section of $ K^+$-meson production
for $p + ^{12}C$ as a function of the beam energy $T_0$. The solid
and dashed lines show the calculations with the total
spectral function for the two-step and direct reaction
mechanism, respectively, employing the spectral
function  from~\protect\cite{Sick}. The dotted and
dashed-dotted lines show
the calculations with  mean-field approach (dashed curve in Fig.~3)
for the two-step
and direct reaction mechanism, respectively, employing a quasifree
dispersion relation.
The dots show the experimental data from~\protect\cite{Koptev}.
The solid and the dotted lines were calculated with the same
elementary cross sections.}
\end{figure}

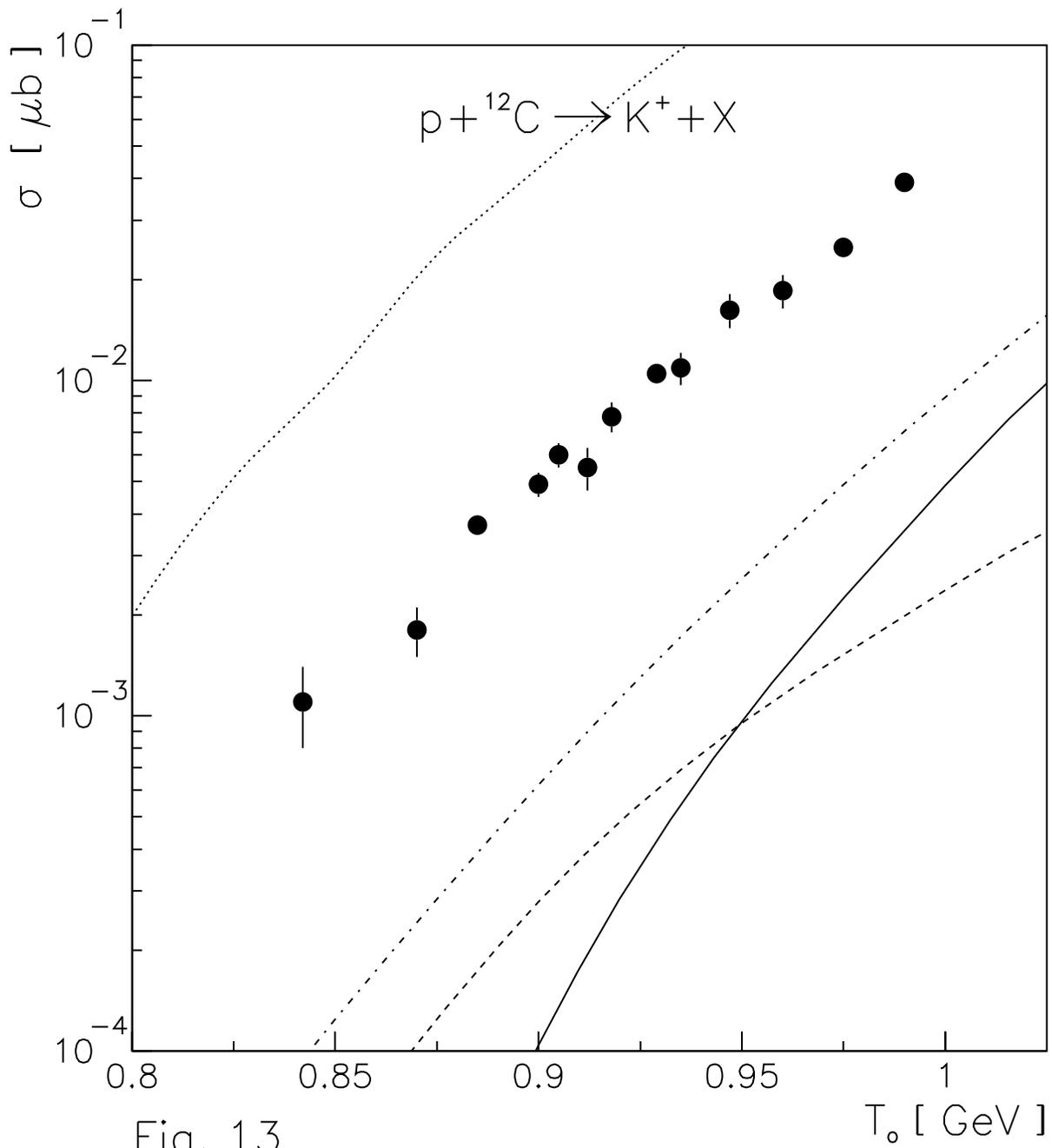
\begin{figure}
\caption{\label{e6ka1}Cross section of $ K^+$-meson production
for $p + ^{12}C$ as a function of the beam energy $T_0$. The solid
and dashed lines are results using only the correlated part
of the spectral function from~\protect\cite{Atti1}
for the two-step and direct reaction
mechanism, respectively. The dotted and dashed-dotted lines show
the on-shell calculations for the two-step
and direct reaction mechanism, respectively.
Experimental data are from~\protect\cite{Koptev}.}
\end{figure}

\begin{figure}
\caption{\label{e6h}Cross section for $\rho$ , $\omega$
and $\phi$-meson production
in $p + ^{12}C$ reactions as a function of the beam energy $T_0$. The solid
and dashed lines show the calculations with the total
spectral function~\protect\cite{Sick} for the two-step and direct reaction
mechanism, respectively.} 
\end{figure}

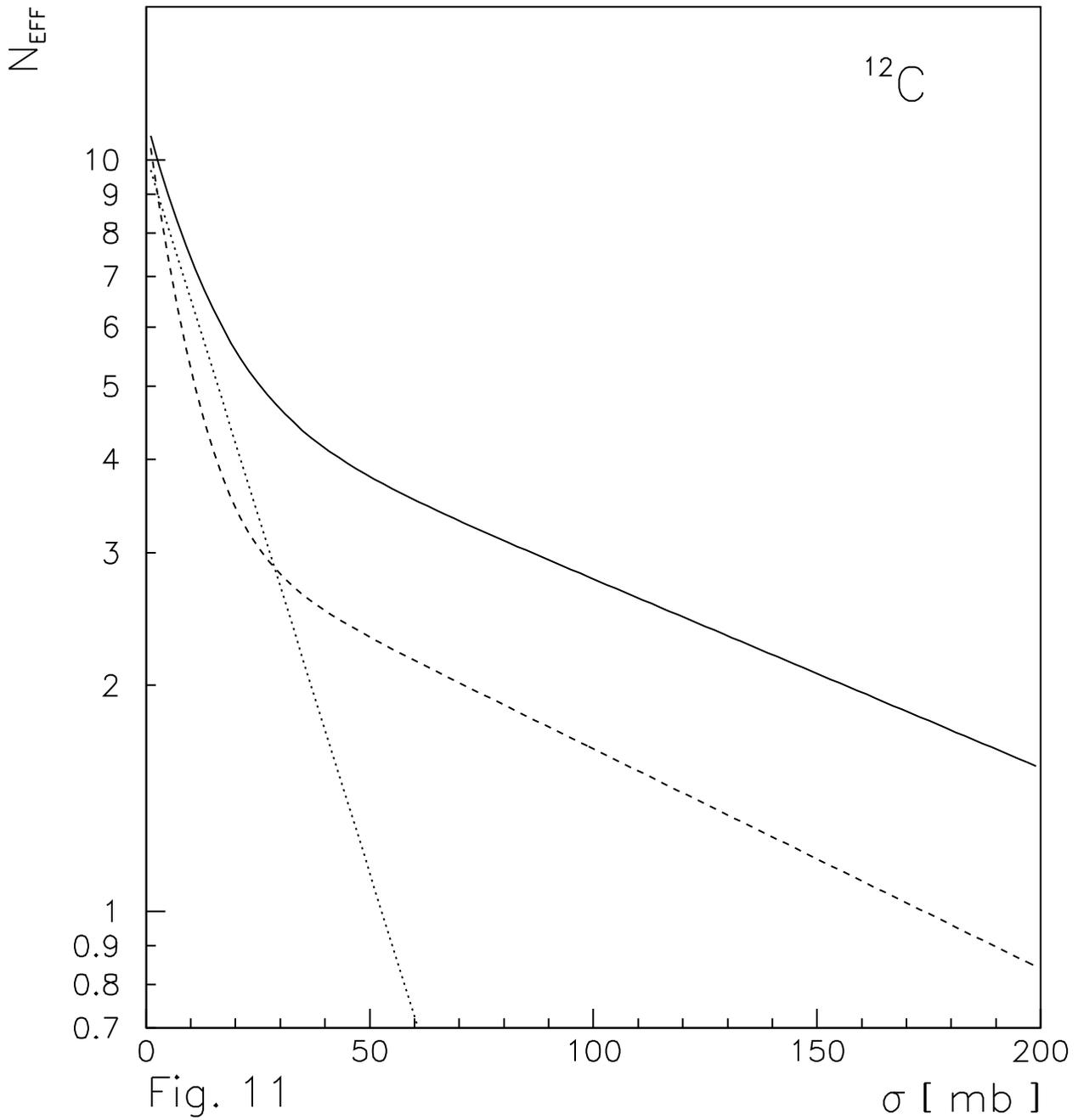
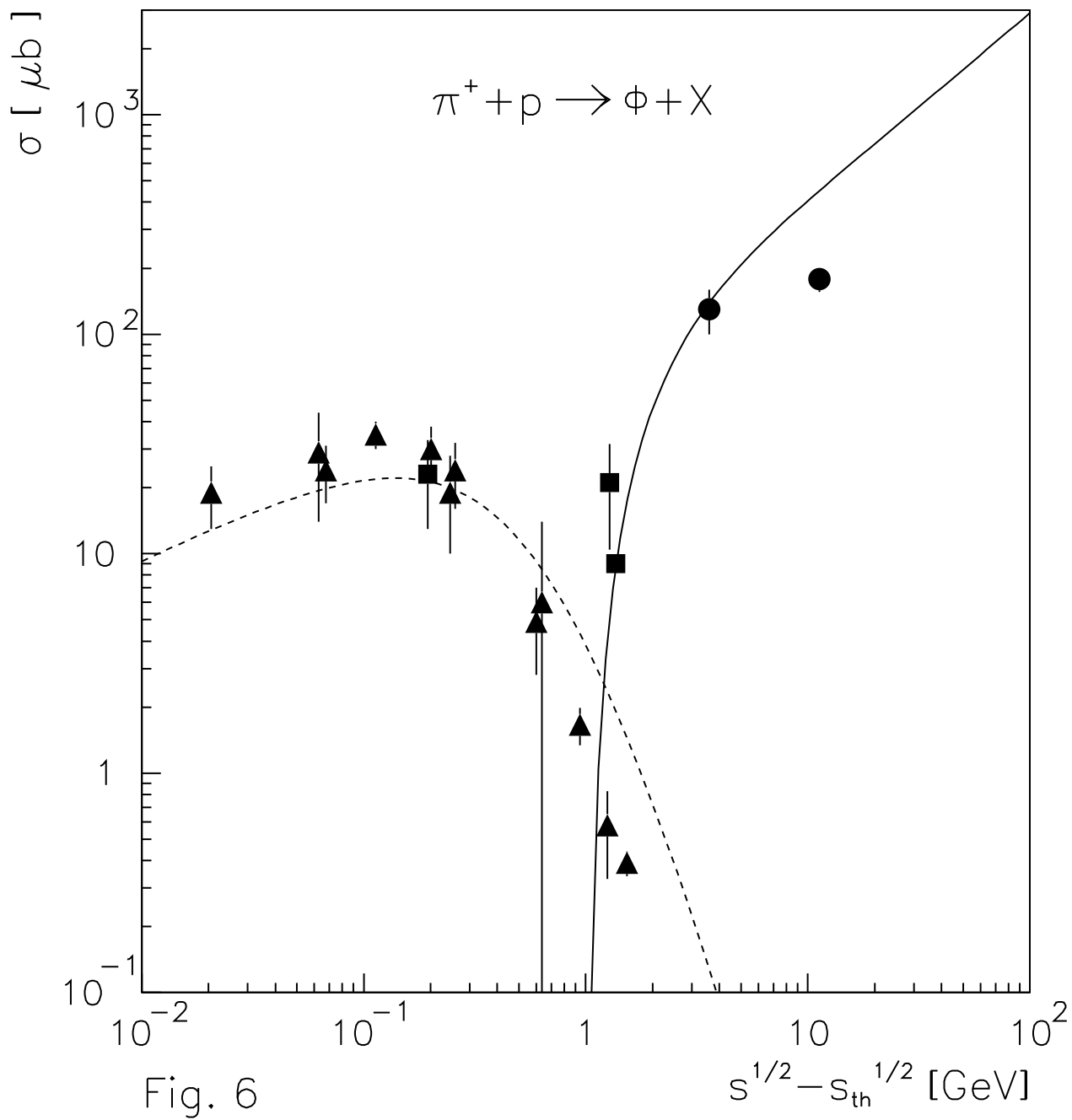
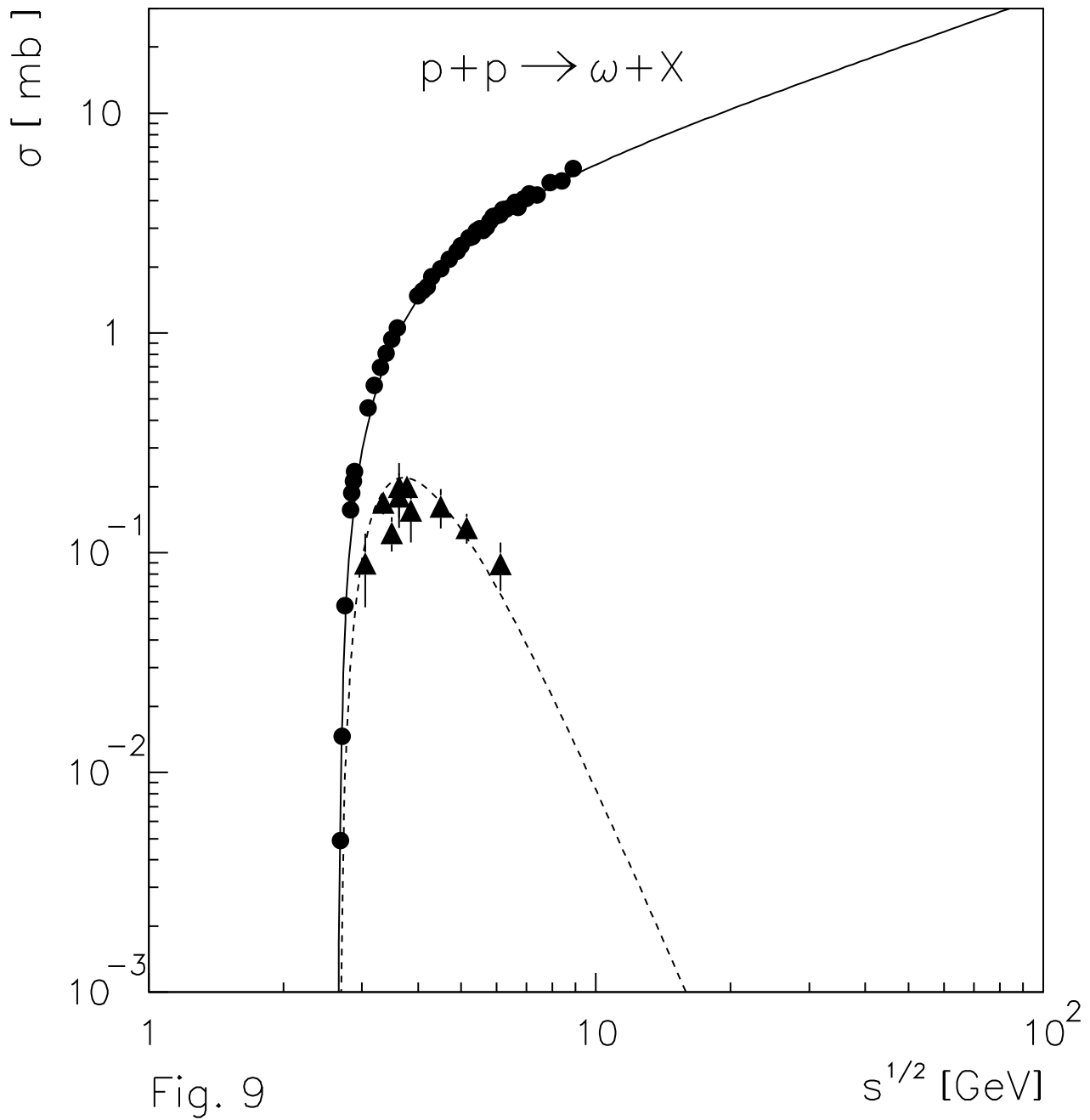
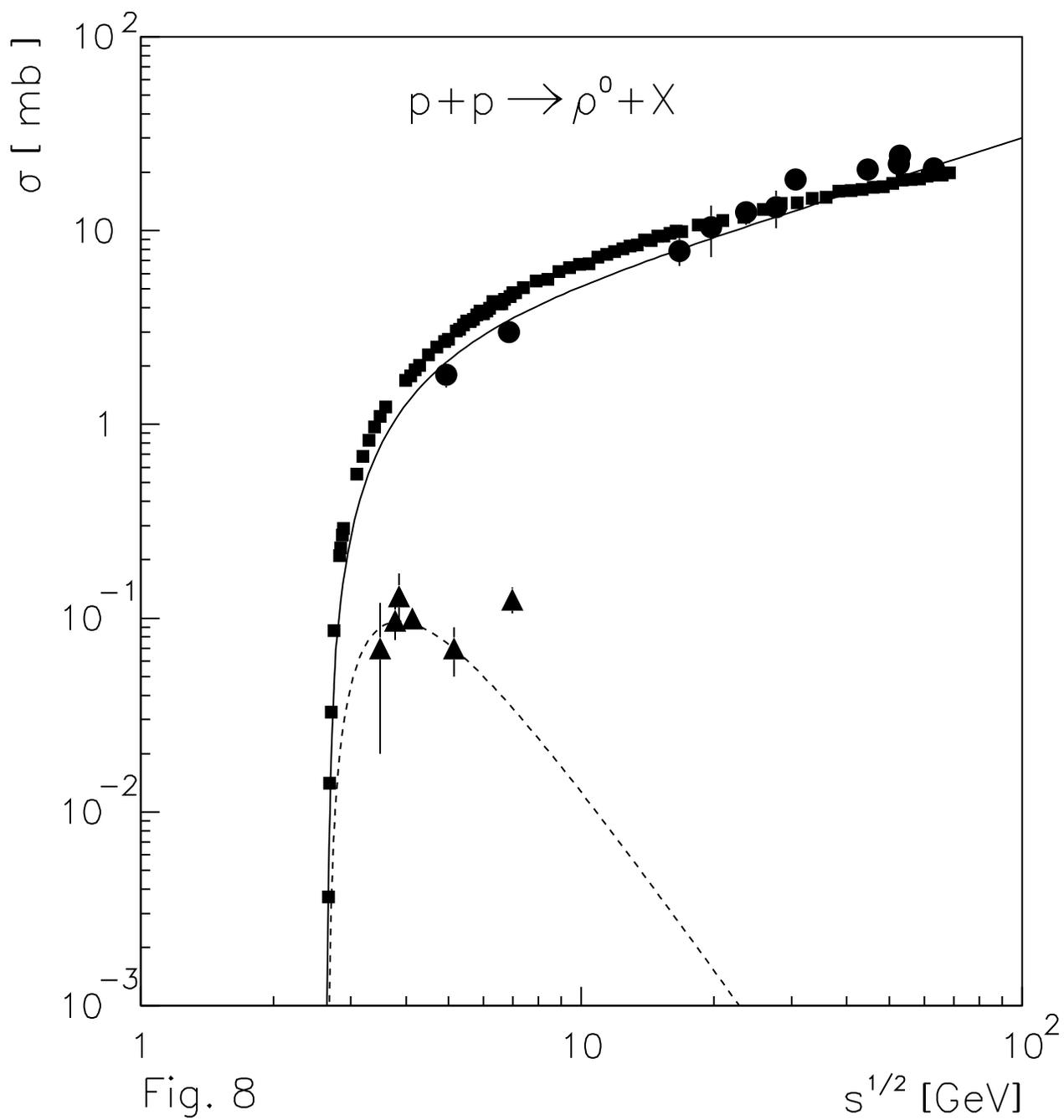
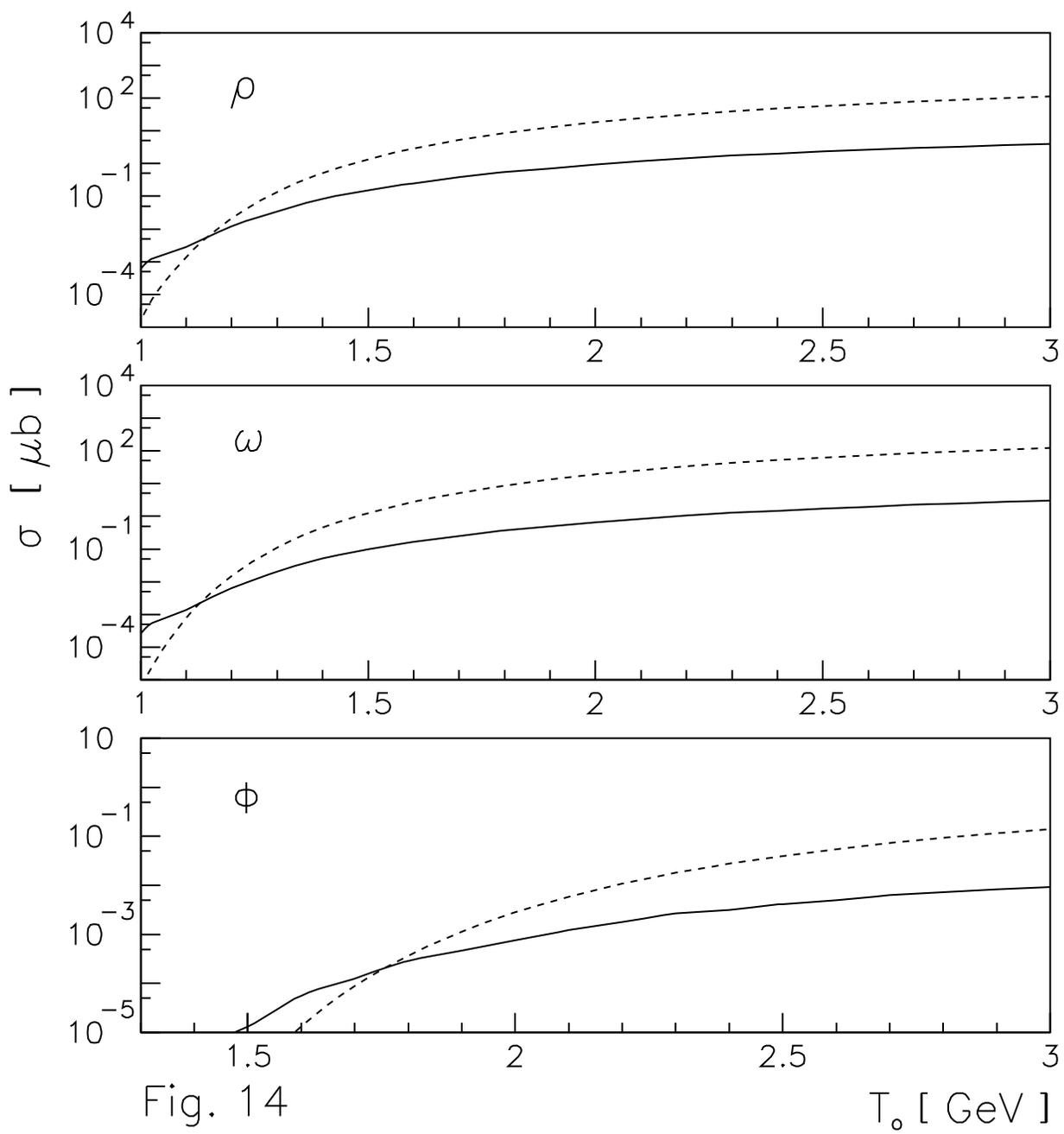
\end{document}